\documentclass[%
 reprint,
 amsmath,amssymb,
 aps,
]{revtex4-2}

\usepackage{amsmath,amssymb,amsthm}
\usepackage{physics}
\usepackage{hyperref}
\usepackage{graphicx}
\usepackage{dcolumn}
\usepackage{bm}
\usepackage{cancel}
\usepackage{amsfonts}
\usepackage{mathrsfs}
\usepackage{mathtools}
\usepackage{xcolor}
\usepackage{comment}
\usepackage{float}
\usepackage{enumitem}

\def\dj{d\kern-0.4em\char"16\kern-0.1em}
\def\Dj{\mbox{\raise0.3ex\hbox{-}\kern-0.4em D}}
\def\kbar{{\mathchar'26\mkern-9muk}}

\begin{document}

\title{Twisted holographic superconductors in external magnetic field}

\author{Jovan Potrebi\'{c}}
\email{jovan.potrebic@ff.bg.ac.rs}
\author{Dragoljub Go\v{c}anin}
\email{dragoljub.gocanin@ff.bg.ac.rs}
\affiliation{%
Faculty of Physics, University of Belgrade\\ Studentski Trg 12-16, 11000 Belgrade, Serbia
}

\def\kbar{{\mathchar'26\mkern-9muk}}

\begin{abstract}

Among the various applications of the AdS/CFT correspondence in condensed matter physics, the realization of the phase transition between the normal and superconducting phases in holographic quantum field theory is of particular importance. Following seminal papers on holographic superconductors that introduced the basic framework, one major line of development has focused on capturing the Meissner effect with all relevant parameters, which requires the inclusion of an external magnetic field. Although a complete holographic description of a superconductor is still lacking, the basic elements of the gravitational systems dual to what can most accurately be characterized as a charged superfluid have been established. Using holographic setups to describe three- and four-dimensional superconductors, we investigate the effect of noncommutative twist deformation of bulk fields on the phase transition parameters, such as the critical magnetic field. In a broader context, our results represent the first systematic attempt to elucidate the role of noncommutative gauge field theory as part of the bulk description of condensed matter systems.

\end{abstract}

\maketitle

\section{Introduction}
\label{sec:Introduction}

\twocolumngrid
One of the most significant discoveries in string theory that has provided a fruitful ground for future investigations – many of which are independent of string theory itself – is the AdS/CFT correspondence \cite{Maldacena:1997re,Gubser:1998bc,Witten:1998qj}. This conjectured duality states that string theory in (asymptotically) $(d+1)$-dimensional anti-de Sitter ($AdS_{d+1}$) spacetime has an equivalent holographic description in terms of a conformal field theory (CFT) defined, in a certain sense, on its $d$-dimensional boundary. Apart from being a way of defining quantum gravity, holographic duality can be used as a tool for studying various aspects of quantum field theories (QFTs) in gravitational terms. Of particular interest is the limit in which strongly coupled QFTs on the boundary correspond to classical gravitational systems in the bulk, which allows us to study various condensed matter systems in a bottom-up approach. 

As shown in \cite{Gubser:2008px}, spontaneous $U(1)$ symmetry breaking in bulk geometries with a horizon can be used to describe the transition from the normal to the superconducting phase in the dual QFT. This observation led to the seminal work of Hartnoll \textit{et al}. \cite{Hartnoll:2008kx,Hartnoll:2008vx,Herzog2009}, in which a model of a holographic superconductor was first introduced. Over time, many extensions of this model and various applications of holographic duality to condensed matter physics have appeared \cite{Horowitz:2008bn,Nakano:2008x,Amado:2009ts,Koutsoumbas:2009pa,Maeda:2008ir,Sonner:2009tf,Zeng:2009hv,Pang:2009pd,Cai:2009hn,Siopsis:2010uq}.
One of the first questions that arose was how to extend the model to incorporate magnetic fields. A realistic holographic model is expected to exhibit the Meissner effect and to allow for the computation of the superconducting coherence length and magnetic penetration depth. To capture these phenomena, dynamical gauge fields on the boundary theory are required. However, the original model of Hartnoll \textit{et al.} considered $U(1)$ gauge fields in the bulk, which, according to the holographic dictionary, correspond to a global $U(1)$ symmetry on the boundary. Consequently, although this construction was referred to as an AdS/CFT superconductor, it was more accurately a holographic description of a ''charged superfluid. The first attempts`` to incorporate constant magnetic fields appeared shortly thereafter \cite{Nakano:2008x,Albash:2008bm,Hartnoll2009}.

The appearance of a phase transition in the $(2+1)$-dimensional case is particularly intriguing, since the Mermin-Wagner-Hohenberg theorem \cite{MerminWagner1966,Hohenberg1967} forbids spontaneous symmetry breaking due to large fluctuations in low-dimensional systems. However, the holographic realization of superconductivity is fundamentally different. In this setup, the phase transition becomes possible due to the peculiarities of the large-$N$ limit, which effectively suppresses fluctuations and thereby allows algebraic long-range order to survive. The question of condensation in higher dimensions was soon addressed in \cite{Gregory:2009fj}. Shortly after, models of $(3+1)$-dimensional holographic superconductors were constructed with \cite{Ge:2010aa} and without magnetic fields \cite{Gregory:2009fj,Pan:2009xa,Pan:2010px}. Related studies have addressed flux periodicity and quantum hair in holographic superconductors, magnetic effects in the insulator/superconductor transition, holographic superfluids/superconductors in dilaton-gravity backgrounds, emergent gauge fields, and phase transitions in dilaton holography with global or local symmetries \cite{Montull:2011im, Montull:2012fy, Salvio:2012at, Domenech:2010nf, Salvio:2013iaa}.

On the other hand, background magnetic fields (or their generalizations) are known to be linked, in certain setups, with the emergence of the structure of noncommutative (NC) geometry. A typical example is provided by an effective description of a single Landau level (usually the lowest Landau level) as a NC plane generated by noncommuting guiding-center coordinates of the electron's 
cyclotron motion in a strong orthogonal magnetic field, $[R_{x},R_{y}]=i\ell_{B}^{2}$, with the NC scale set by the magnetic length $\ell_{B}=\sqrt{\tfrac{\hbar}{eB}}$, see \cite{Girvin1986GMP}. A similar structure appears in the seminal paper by Dong and Senthil \cite{DongSenthil2020NCFieldTheoryCFL} where composite fermion fields on NC space are used to describe fluctuations in the low-energy regime; see \cite{GocaninEtAl2021DiracCFLNoncomm}. 
As a kind of generalization, in the string theory setting, Seiberg and Witten found that, within an appropriate limit, the end points of an open string attached to a D-brane immersed in a background Kalb-Ramond B field effectively probe a NC space, with the NC scale inversely proportional to the B field \cite{SeibergWitten1999NC}; see also \cite{ArdalanArfaeiSheikhJabbari1999NCGeom, ChuHo1999NCOpenString}. The effective description of the brane is given by NC gauge field theory.

Motivated by the general expectation that the classical structure of spacetime breaks down in some way at sufficiently small length scales (presumably Planckian), one can promote spacetime coordinates to operators that satisfy nontrivial commutation relations. The most common of these are the canonical, or $\theta$-constant, commutation relations, $[\hat{x}^{\mu},\hat{x}^{\nu}]=i\theta^{\mu\nu}$, where the NC parameters $\theta^{\mu\nu}$ form a constant antisymmetric matrix, $\theta^{\mu\nu}=-\theta^{\nu\mu}=\kbar \ell^{2}_{\text{NC}}$, thus introducing a NC geometry of spacetime itself associated with the NC scale $\ell_{\text{NC}}$; here, $\kbar\in\mathbb{R}$ is a dimensionless deformation parameter. In a NC spacetime, the uncertainty relations $\Delta x^{\mu}\Delta x^{\nu}\geq\frac{1}{2}\vert\theta^{\mu\nu}\vert$ make ideal localization of events (spacetime points) impossible.
An interesting aspect of having a (canonical) NC spacetime as a background is that it can deform the Landau levels of an electron in a magnetic field by effectively changing the value of the magnetic field, $B_{\text{eff}}=B(1+B\theta)$. Note that spacetime noncommutativity alone is not sufficient to replace the external magnetic field; it can only modify an already existing one \cite{DimitrijevicCiric:2018so23NCED}.   
In this work, we build on the framework developed in \cite{Albash:2008bm,Ge:2010aa} and investigate the consequences of having NC fields in the bulk on the physics of $(2+1)$- and $(3+1)$-dimensional holographic superconductors in the presence of external magnetic fields. We show that NC gauge field structure in the bulk has a direct effect on the parameters of the dual boundary system. This is a first step in a systematic investigation of the potential that NC field theory has for capturing the physics of superconductors in the setting of holographic duality. 

The paper is organized as follows. In Sec. \ref{sec:NC}, we introduce some basic elements of NC field theory to set the stage. In Sec. \ref{sec:3D}, we present a NC model of a $(2+1)$-dimensional holographic superconductor in a homogeneous external magnetic field and use numerical methods to determine the effects of noncommutativity on the condensation parameters. In Sec. \ref{sec:4D}, we use a numerical and analytical approach to study a model of a NC $(3+1)$-dimensional holographic superconductor. Conclusions and outlook make up the final section. The details of the calculations are presented in Appendixes \ref{appendixA} - \ref{appendixC}.

\section{NC Field Theory}
\label{sec:NC}

The models of holographic superconductors that we are going to use in the following sections involve ordinary dual gravitational descriptions in terms of classical black hole geometries with AdS asymptotics. What we will be interested in is what happens with a holographic superconductor (its critical magnetic field and condensate) when we apply a NC twist of the bulk fields that deforms the classical bulk. Note that in our approach, the NC structure does not pertain to the boundary system, which is, in principle, considered to be some realistic superconductor accessible to observations, but merely to its dual geometric description. In this sense, NC geometry is used as a tool that could potentially broaden the holographic language and perhaps capture some new aspect of a lower-dimensional condensed matter theory system.     

In general, an (Abelian) twist operator can be defined in a coordinate-free fashion using a set $\{X_I\, \vert\, I=1,\dots,s\leq D\}$ of mutually commuting vector fields (note that their number need not be equal to the number of spacetime dimensions $D$) and a constant real antisymmetric matrix $\theta^{IJ}$, as
\begin{equation}
  \mathcal{F}
  =\exp\left(-\frac{i}{2}\theta^{IJ}X_I\otimes X_J\right).
\end{equation}
The twist operator deforms the ordinary commutative product of functions on spacetime into a NC $\star$-product 
\begin{align}\label{star}
 &f_{1}\star f_{2} = \mu\left(\mathcal{F}^{-1}
    (f_{1}\otimes f_{2})\right)\nonumber\\
    &=f_{1}f_{2}+\frac{i}{2}\theta^{IJ}X_{I}[f_{1}]X_{J}[f_{2}]+\mathcal{O}(\theta^{2}),
\end{align}
where $\mu\left(f_{1}\otimes f_{2}\right)=f_{1}\cdot f_{2}$ is the multiplication map and $X_{I}[f]=X^{\mu}_{I}\partial_{\mu}f$.
Since the vector fields \(X_I\) commute, the \(\star\)-product is associative. The system of commuting vector fields $\{X_{I}\}$ that generate the twist operator
are an independent structure on the spacetime manifold that provides a generally covariant formulation of a NC theory. However, there is no universal criterion for determining what set of vector fields we should choose in a particular situation. We can always choose (at least locally) a system of coordinates $x^{\mu}$ adapted to $\{X_{I}\}$ by setting $X_{I}=\partial_{\mu}$ so that (\ref{star}) reduces to the standard Moyal-Weyl-Gronewold $\star$-product
\begin{align}\label{Moyal}
f_{1}\star f_{2}
&= f_{1}f_{2}+\frac{i}{2}\theta^{\mu\nu}\partial_{\mu}f_{1}\partial_{\nu}f_{2}+\mathcal{O}(\theta^{2}).
\end{align}
When applied on coordinate functions themselves, it gives us the $\theta$-constant $\star$-commutator relations,
$[x^{\mu},x^{\nu}]_{\star}=x^{\mu}\star x^{\nu}-x^{\nu}\star x^{\mu}=i\theta^{\mu\nu}=const.$

In the absence of a universal criteria, it seems reasonable to base our choice on the isometries of spacetime. Taking a set of commuting Killing vector fields $\{K_{I}\}$ of a given spacetime as generators of an Abelian twist ensures that, while the algebraic structure of the theory gets NC deformed, the coupling with the metric and all the objects built out of it (connection, curvature) remains unchanged. 
Namely, by choosing the coordinate system 
$y^{\alpha}$ such that $K_{I}=\delta^{\alpha}_{I}\tfrac{\partial}{\partial y^{\alpha}}\equiv \delta^{\alpha}_{I}\partial_{\alpha}$,  the Killing equation yields $\partial_{\alpha}g_{\mu\nu}(y)=0$, ensuring that the coupling with the metric remains commutative as in the classical theory, 
\begin{align}g_{\mu\nu}\star(\cdots)&=g_{\mu\nu}\cdot(\cdots)+\frac{i}{2}\theta^{IJ}\delta^{\alpha}_{I}\partial_{\alpha}g_{\mu\nu}(y)\mathcal{L}_{K_{J}}(\cdots)\nonumber\\
&=g_{\mu\nu}\cdot(\cdots).
\end{align} 
While it leaves the coupling with spacetime geometry unchanged, a Killing twist can result in NC-deformed equations of motion for matter fields.

\subsection{Seiberg-Witten gauge field theory}
\label{subsec:SW_gauge_QFT}

To define an action for NC gauge field theory we first replace the ordinary commutative product by a NC $\star$-product in the classical action. But to maintain gauge invariance of the action, one also has to introduce NC fields (denoted by a hat symbol) that transform in representations of a NC-deformed symmetry group. In the Seiberg-Witten (SW) construction, a variation of a NC matter field (in the fundamental representation) under an infinitesimal NC gauge transformation is given by
\begin{equation}
\delta_{\hat{\epsilon}}^\star \, \hat{\Phi} 
= i \, \hat{\epsilon} \star \hat{\Phi}, 
\label{eq:NC22}
\end{equation}
where $\hat{\epsilon}(x)$ is a NC gauge parameter. However, if we assume $\hat{\epsilon}$ to be Lie algebra valued, NC gauge transformations will fail to satisfy the closure axiom. The problem is resolved by taking $\hat{\epsilon}$ from the universal enveloping algebra (UEA) of the starting (undeformed) Lie algebra, in which case we have a closure 
\begin{equation}
[\delta_{\hat{\epsilon}_1}^\star, \delta_{\hat{\epsilon}_2}^\star]_{\star}
= \delta^{\star}_{-i [\hat{\epsilon}_1 , \hat{\epsilon}_2]_{\star}}.
\label{eq:NC23}
\end{equation}
In the SW construction, NC covariant derivative and NC field strength inherit the structure of their classical counterparts,  
\begin{equation}
\hat{D}_\mu \hat{\Phi} = \partial_\mu \hat{\Phi} - i \, \hat{A}_\mu \star \hat{\Phi}, 
\label{eq:NC24}
\end{equation}
\begin{equation}
\hat{F}_{\mu\nu} 
= \partial_\mu \hat{A}_\nu - \partial_\nu \hat{A}_\mu 
- i [\hat{A}_\mu , \hat{A}_\nu]_{\star}.
\label{eq:NC27}
\end{equation}
Note that even if we work with an Abelian gauge theory with $U(1)$ gauge group, the NC $U(1)_{\star}$ field strength still has the non-Abelian character due to $\star$-product. However, the fact that $\hat{F}_{\mu\nu}$ is also UEA valued means that the system has infinitely many new degrees of freedom. To eliminate these unwanted degrees of freedom we use the SW map \cite{SeibergWitten1999NC}, which allows us to express the NC degrees of freedom in terms of the original commutative ones. The basic principle is that NC gauge transformations are induced by the commutative ones, 
\begin{equation}
\delta^{\star}_{\hat{\epsilon}} \hat{A}_\mu(A) = \hat{A}_\mu(A + \delta_{\epsilon} A) - \hat{A}_\mu(A),
\label{eq:NC27}
\end{equation}
where $\delta_{\epsilon} A_\mu = -\partial_\mu \epsilon$ and $\delta^{\star}_{\hat{\epsilon}} \hat{A}_\mu = -\partial_\mu \hat{\epsilon} + i  [\hat{\epsilon},\hat{A}_\mu]_{\star}$. Now we can solve the differential equation \ref{eq:NC27} perturbatively and derive the SW representation of the NC field $\hat{A}_\mu$ as a power series in $\theta^{\mu\nu}$ with coefficients built out of classical fields (likewise for $\hat{\Phi}$ and $\hat{F}_{\mu\nu}$),
\begin{align}
\hat{A}_\mu &= A_\mu 
- \frac{1}{4} \, \theta^{\alpha\beta} \{ A_\alpha, \partial_\beta A_\mu + F_{\beta\mu} \} 
+ \mathcal{O}(\theta^2), 
\label{eq:NC29} \\
\hat{F}_{\mu\nu} &= F_{\mu\nu} 
- \frac{1}{4} \, \theta^{\alpha\beta} \{ A_\alpha, (\partial_\beta + D_\beta) F_{\mu\nu} \} \nonumber\\
&+ \frac{1}{2} \, \theta^{\alpha\beta} \{ F_{\alpha\mu}, F_{\beta\nu} \} 
+ \mathcal{O}(\theta^2), 
\label{eq:NC30} \\
\hat{\Phi} &= \Phi 
- \frac{1}{4} \, \theta^{\alpha\beta} A_\alpha (\partial_\beta + D_\beta) \Phi 
+ \mathcal{O}(\theta^2).
\label{eq:NC31}
\end{align} 
Finally, the NC action can be organized as a perturbative expansion in the powers of $\theta^{\mu\nu}$, maintaining the (undeformed) gauge invariance at each order.  

\section{$3$D Twisted Holographic Superconductor in an External Magnetic Field}

\label{sec:3D}

A model of a $(2+1)$-dimensional holographic superconductor immersed in an external magnetic field is constructed in \cite{Albash:2008bm} and the bulk action is given by  
\begin{align} \label{eq:4D1}
&S=\frac{1}{2\kappa_4^2}\int d^4x\sqrt{-g}\Bigg( R+\frac{6}{L^2}\nonumber\\ &+L^2\left[-\frac{1}{4}F_{\mu\nu}F^{\mu \nu} -(D_{\mu}\Phi)^{*}D^{\mu}\Phi-V(|\Phi|)\Bigg]\right),
\end{align}
where $D_{\mu}$ is the $U(1)$-covariant derivative that involves (when appropriate) the standard symmetric and metric-compatible Christoffel connection $\Gamma_{\mu}$ of the curved background, and we choose the quadratic potential $V(|\Phi|)=m^2|\Phi|^2$ (which turns out to be enough) with $m^2=-2/L^2$ safely above Breitlohner-Freedman bound \cite{Breitenlohner:1982jf}. Moreover, we will work in the probe limit in which Einstein-Maxwell sector decouples from the scalar, and use a self-consistent, fully backreacted, electrically and magnetically charged, i.e., dyonic Reissner-Nordström (RN) AdS black hole as a fixed background for the dynamical scalar field. This background is a particular (exact) solution of the Einstein-Maxwell field equations with the metric given by
\begin{align}\label{eq:4D2}
ds^2 &= \frac{L^2 \alpha^2}{z^2} \left( -f(z)\, dt^2 + dx^2 + dy^2+\frac{dz^2}{\alpha^{2} f(z)} \right),
\end{align}
where 
\begin{align}\label{eq:4D4}
f(z) &= 1 + (h^2 + q^2) z^4 - \big(1 + h^2 + q^2\big) z^3 \nonumber\\ 
      &=  (1-z) \big( z^2 + z + 1 - (h^2 + q^2) z^{3} \big),
\end{align}
and the gauge field strength is 
\begin{equation}
F = 2 h \alpha^2 \, dx \wedge dy + 2 q \alpha \, dz \wedge dt.
\label{eq:4D3}    
\end{equation}
In this coordinate system, $z$ is a dimensionless radial coordinate, scaled such that the black hole horizon is located at $z_h=1$ and the asymptotic boundary at $z=0$. The parameters $\alpha$, $q$, and $h$ are related to the mass, electric charge, and magnetic charge of the black hole, respectively, with $\alpha$ being the only dimensionful parameter among them, having dimensions of inverse length. In dual boundary theory, these parameters determine the temperature, magnetic field, and charge density. The only remaining dimensionful parameter is $L$, the AdS radius.
The temperature of the black hole and its dual field theory is obtained via Bekenstein-Hawking formula \cite{Gibbons:1979},
\begin{equation}
    T=\frac{1}{\beta}=\frac{\alpha}{4\pi}(3-h^2-q^2),
    \label{eq:4D5}
\end{equation}
with $h^2+q^2\leq3$ to ensure positivity. The case when this inequality is saturated corresponds to an extremal, zero-temperature black hole.
Since we are interested in axially symmetric configurations, we choose a gauge adapted to these symmetries, namely
\begin{equation}
    A =  h \alpha^2 (x \, dy-y\, dx) + 2 q \alpha (z - 1) \, dt.
    \label{eq:4D6}
\end{equation}
Following \cite{Kobayashi:2006sb}, we include the term $-2q\alpha$ so that the gauge potential remains regular at the horizon. This gauge potential produces a constant magnetic field on the boundary theory~\cite{Hartnoll:2007ai},
\begin{equation}
    B=F_{xy}=2h\alpha^2.
    \label{eq:4D7}
\end{equation}
Now that we have a well-defined classical (undeformed) bulk configuration, we want to introduce a NC twist deformation, adapted to the isometries of the bulk metric, and see how this affects the boundary parameters. Applying the procedure described in Sec. \ref{sec:NC} on the classical action \eqref{eq:4D1}, we obtain its NC extension,
\begin{align}
S_{\text{NC}} =& \frac{1}{2\kappa_4^2} \int d^4 x  \sqrt{-g}  \bigg(
R + \frac{6}{L^2} \nonumber\\
+& L^2 \bigg[
- \frac{1}{4} g^{\mu \alpha}  g^{\nu \beta}  \hat{F}_{\mu \nu} \star \hat{F}_{\alpha \beta}\nonumber \\
-& g^{\mu \nu}  (D_\mu \hat{\Phi})^* \star D_\nu \hat{\Phi} 
-m^2\hat{\Phi}^*\star\hat{\Phi} 
\bigg] 
\bigg),
\label{eq:4D10}
\end{align}
where $D_\mu \hat{\Phi} = \partial_\mu \hat{\Phi} - i  \hat{A_\mu} \star \hat{\Phi}$ is a $U(1)_\star$ covariant derivative of the $U(1)_\star$ scalar field $\hat{\Phi}$, while $\hat{A}_\mu$  denotes a $U(1)_\star$ gauge field. It is important to note that the spacetime metric is not promoted to a NC field, and that it couples classically (with ordinary product) to other fields. This is the consequence of using a Killing twist, which leaves the background metric fixed and only deforms the $U(1)$  algebra of gauge transformations. The NC action (\ref{eq:4D10}) is invariant under twisted $U(1)_{\star}$ gauge transformations. 
  From the form of the dyonic RN-AdS black hole's metric~\eqref{eq:4D2} and \eqref{eq:4D4}, we can readily identify the Killing vectors: $\{\partial_t, \partial_x, \partial_y, x\partial_y - y\partial_x\}$, which close the $\mathbb{R} \oplus ISO(2)$ algebra. Since we are only interested in an Abelian Killing twist, possible twist generators reduce to $\{\partial_t, \partial_x, \partial_y\}$. For our purposes, however, we restrict to a twist generated by $\partial_x$ and $\partial_y$, namely,
\begin{equation}
    \mathcal{F} = \exp\left[-\frac{i \kbar}{2\alpha^2}\left(\partial_x \otimes \partial_y - \partial_y \otimes \partial_x\right)\right],
    \label{eq:4D9}
\end{equation}
where $\theta^{\mu\nu} = \theta^{IJ} \, \delta^\mu_I \, \delta^\nu_J$ such that $\theta^{xy} = -\theta^{yx} \equiv \frac{\kbar}{\alpha^2}$, with all other components vanishing. 

Using the SW representation (\ref{eq:NC29}) and (\ref{eq:NC31}) of the NC fields $\hat{A}_{\mu}$ and $\hat{\Phi}$ as perturbative expansions in powers of $\alpha^2\theta^{\mu\nu} \sim \kbar$ around the classical fields, the NC action (\ref{eq:4D10}) can be expanded up to first order in $\kbar$, $S_{\text{NC}}=S^{(0)}+S^{(1)}+\mathcal
{O}(\kbar^{2})$, where $S^{(0)}$ is the classical (undeformed) action (\ref{eq:4D1}) and the $\kbar$-linear correction 
\begin{align}\label{eq:4D12}
&S^{(1)}=\frac{\theta^{\alpha \beta}L^2}{2\kappa_4^2}\int d^4x\sqrt{-g}\nonumber\\
&\times\Bigg(
\frac{1}{8}F_{\alpha \beta}F_{\mu \nu} F^{\mu \nu}-\frac{1}{2}F_{\mu \alpha} F_{\nu \beta} F^{\mu \nu}  +\frac{m^2}{4} F_{\alpha \beta} \vert\Phi\vert^{2} \nonumber\\
&+    g^{\mu\nu} \Bigg[
 \frac{1}{4}(D_\mu \Phi)^*  D_\nu \Phi F_{\alpha\beta} 
- \frac{1}{2}(D_\mu \Phi)^* F_{\alpha\nu} D_\beta \Phi\nonumber\\
&\hspace{1cm}- \frac{1}{2}(D_\beta \Phi)^* F_{\alpha\nu} D_\mu \Phi\Bigg] \Bigg).
\end{align}
The key equations for expanding the products of NC fields are given in Appendix \ref{appendixA}.
Note that there are no new fields in the NC action, just new interaction terms composed out of the original fields. Everything is still minimally coupled to gravity. This NC action was also studied in \cite{DimitrijevicCiric:2019vtx} for a massless scalar.    

By varying the NC action with respect to $\Phi^{*}$, we obtain the NC equations of motion for the scalar field,
\begin{align}
&\Big(
D^\mu D_\mu 
- m^2\Big) \Phi
+  
 \theta^{\alpha\beta} 
\Bigg(\frac{m^2}{4} F_{\alpha\beta} \Phi
-\frac{1}{4}D_{\mu}(F_{\alpha\beta} D^\mu \Phi)\nonumber\\
&+\frac{1}{2}D_{\mu}(F_{\alpha} {}^{\mu}D_{\beta}\Phi)+\frac{1}{2}D_{\beta}(F_{\alpha\mu}D^{\mu}\Phi) 
\Bigg)=0,
\end{align}
or, making the spacetime connection explicit,
\begin{align}\label{eq:4D13}
&\Big(
D^\mu D_\mu 
- m^2\Big) \Phi
+ \frac{m^2}{4}\theta^{\alpha\beta}  F_{\alpha\beta} \Phi\nonumber\\
&-\frac{g^{\mu\nu}}{4}\theta^{\alpha\beta}\Big((\partial_{\mu}-iA_{\mu})(F_{\alpha\beta}D_{\nu}\Phi)-\Gamma_{\mu\nu}^{\lambda}F_{\alpha\beta}D_{\lambda}\Phi \Big)\nonumber\\
&+\frac{g^{\mu\nu}}{2}\theta^{\alpha\beta}\Big((\partial_{\mu}-iA_{\mu})(F_{\alpha\nu}D_{\beta}\Phi)-\Gamma_{\mu\nu}^{\lambda}F_{\alpha\lambda}D_{\beta}\Phi \Big)\nonumber\\
&+\frac{g^{\mu\nu}}{2}\theta^{\alpha\beta}(\partial_{\beta}-iA_{\beta})(F_{\alpha\mu}D_{\nu}\Phi)=0.
\end{align}
On the other hand, by varying the NC action with respect to $A_{\mu}$, we come to the NC equations of motion for the gauge field. 
It turns out that, in the probe limit, when the scalar decouples and we only consider the Einstein-Maxwell sector of the theory, our classical dyonic RN-AdS black hole described by \eqref{eq:4D2} and \eqref{eq:4D3} remains an exact solution of the NC Maxwell field equations in curved spacetime because, when evaluated on this solution, the NC correction exactly vanishes; see Appendix \ref{appendixB} for details. Thus, it is consistent to keep the same background and consider only the NC field equations for the scalar field even after the twist. 

Since we are interested in axially symmetric field configurations, it is convenient to adopt polar coordinates $(x,y) \to (r,\phi)$. In these coordinates, the NC equation of motion for $\Phi(r,\phi,z)$ can be explicitly written as
\begin{align}\label{eq:4D16}
&z \Bigg[ z (1- h \kbar) f'(z) \frac{\partial \Phi}{\partial z} + f(z)  (h \kbar-1) \Big( 2 \frac{\partial \Phi}{\partial z} - z \frac{\partial^2 \Phi}{\partial z^2} \Big) \nonumber\\
&+ \frac{z ( h \kbar +1)}{\alpha^2} \Bigg( -2 i \alpha^2  h \frac{\partial \Phi}{\partial \phi} + \frac{\partial^2 \Phi}{\partial r^2} + \frac{1}{r} \frac{\partial \Phi}{\partial r} + \frac{1}{r^2} \frac{\partial^2 \Phi}{\partial \phi^2} \Bigg) \Bigg]\nonumber\\
&- \frac{\Phi}{f(z)} \Big[ f(z) \big(  h \kbar (\alpha^2  h^2 z^2 r^2 -  m^2 L^2) + \alpha^2  h^2 z^2 r^2 \nonumber\\
&+ m^2 L^2 \big) + 4  q^2 (z-1)^2 z^2 ( h \kbar-1) \Big]=0.
\end{align}
For axially symmetric solutions $\Phi(r,\phi,z) = \Phi(r,z)$ the imaginary term $-2 i \alpha^2  h \partial_{\phi} \Phi$ vanishes and Eq. \eqref{eq:4D16} becomes purely real. To proceed, we attempt a separation of variables by writing $\Phi(r,z) = R(r) Z(z)$. This reduces the partial differential equation \eqref{eq:4D16} to two ordinary differential equations,
\begin{equation}
    R''(r) + \frac{1}{r} R'(r) - \Big(\frac{Br}{2}\Big)^2 R(r) = -\lambda^2 R(r),
    \label{eq:4D17}
\end{equation}
\begin{align}\label{eq:4D18}
Z''(z) + \left(\frac{f'(z)}{f(z)} - \frac{2}{z} \right) Z'(z)
+ \Bigg(
\frac{4 q^2 (z-1)^2}{f^{2}(z)}
\nonumber\\
- \frac{m^2 L^2}{z^2 f(z)}
- \frac{\lambda^2}{\alpha^2f(z)}(1+ 2h \kbar)
\Bigg) Z(z) = 0.
\end{align}
Let us emphasize several points about these equations.  
First, during the derivation we consistently neglected terms of order $\mathcal{O}(\kbar^2)$ and higher, keeping only the leading NC corrections.  
Second, in the commutative limit $\kbar \to 0$, we recover the standard equations of~\cite{Albash:2008bm,Hartnoll:2008kx}, as expected.  
Third, Eq. \eqref{eq:4D17} admits exact solutions in terms of confluent hypergeometric functions. Remarkably, for $\lambda^2:= B(2n+1)$, $n \in \mathbb{N}_0$, it reduces to the equation of a two-dimensional harmonic oscillator. In this case, the solutions are Hermite functions. Since the lowest mode is expected to condense first and be the most stable after condensation, we focus on $n=0$, which yields
\begin{equation}
    R(r) = \exp\left(-\frac{ B r^2}{4}\right).
    \label{eq:4D19}
\end{equation}
Eq.~\eqref{eq:4D19} shows that the scalar field decays exponentially with $r$, implying a strong localization of the condensate in the $Oxy$ plane as a reminiscence of the Meissner effect and the formation of the so-called superfluid droplets.

After substituting $m^{2}=-2/L^{2}$, equation \eqref{eq:4D18} reads
\begin{align}\label{eq:4D21}
    Z''(z) &+ \Bigg(\frac{f'(z)}{f(z)} - \frac{2}{z} \Bigg) Z'(z) + \Bigg( \frac{4 q^2 (z-1)^2}{f(z)^2} \nonumber\\ &+ \frac{2}{z^2 f(z)} - \frac{\lambda^2}{\alpha^2f(z)}(1+2h \kbar) \Bigg) Z(z) = 0.
\end{align}
In order to formulate appropriate boundary conditions, we analyze Eq. \eqref{eq:4D21} both near the horizon and near the conformal boundary. Before doing so, let us introduce a dimensionless function
\begin{equation}
Z(z) \;\to\; \tilde{Z}(z) = \alpha L^2 Z(z).
\end{equation}
Near the AdS boundary, the Eq. \eqref{eq:4D21} reduces to 
\begin{equation}
 \tilde{Z}''(z)-\frac{2}{z}\tilde{Z}'(z)+\frac{2}{z^2}\tilde{Z}(z)=0,
    \label{Hom}
\end{equation}
which is an Euler-type differential equation with solutions 
\begin{equation}\label{eq:4D22}
\lim_{z\to0}\tilde{Z}(z)=\Psi^{(1)}z+\Psi^{(2)}z^2,
\end{equation}
where $\Psi^{(1)}$ and $\Psi^{(2)}$ are dimensionless constants. Following \cite{Klebanov:1999tb}, both of these modes prove to be normalizable, meaning that neither acts as a source for the other. This forces us to impose one of the two boundary conditions
\begin{equation}
\Psi^{(1)}=0 \quad \text{or} \quad \Psi^{(2)}=0.
    \label{eq:4D23}
\end{equation}
Here, $\Psi^{(1)}$ is proportional to the vacuum expectation value (VEV) of the $\Delta=1$ operator $\langle \mathcal{O}_1 \rangle$ in the dual boundary theory, while $\Psi^{(2)}$ corresponds to the VEV of the $\Delta=2$ operator $\langle \mathcal{O}_2 \rangle$. Since there are no qualitative differences between the $\Delta=1$ and $\Delta=2$ cases, in our computations we restrict ourselves to the $\Delta=2$ case. 

Next, we analyze the near-horizon behavior of \eqref{eq:4D21}, which can be classified into three distinct regimes:
\begin{itemize}[leftmargin=*]
    \item \textbf{Case 1:} $\frac{\lambda^2}{\alpha^2}(1+2h\kbar) < 2$.  
    In this case, the equation of motion reduces to a Bessel equation
    \begin{align}
        \tilde{Z}''(z)&-\frac{1}{1-z}\tilde{Z}'(z)+\frac{a^2}{1-z}\tilde{Z}(z)=0,
    \label{eq:4D24}\\
    a^2&=\frac{2-\frac{\lambda^2}{\alpha^2}(1+2h\kbar)}{3-h^2-q^2}>0,
    \end{align}
    whose solutions are 
    \begin{align}\label{eq:4D25}
    \lim_{z\to1}\tilde{Z}(z)&= \psi_1 J_0\!\left( 2a \sqrt{1-z} \right) 
+ \psi_2 Y_0\!\left( 2a \sqrt{1-z} \right)
\nonumber\\ 
&\approx
\psi_1 + \psi_2 \left( \frac{2}{\pi} \gamma + \frac{1}{\pi} \log \big(a (1-z)\big) \right),        
\end{align}
    where $\gamma$ is the Euler-Mascheroni constant. To avoid divergences, we set $\psi_2=0$ as the boundary condition.

\item \textbf{Case 2:} $\frac{\lambda^2}{\alpha^2}(1+2h\kbar) = 2$.  
    Here, the equation simplifies to 
    \begin{align}
    &\tilde{Z}''(z)-\frac{1}{1-z}\tilde{Z}'(z)+b^2\tilde{Z}(z)=0,
        \label{eq:4D26}\\
    b^2&=\frac{4q^2}{(3-h^2-q^2)^2}+\frac{4}{3-h^2-q^2}>0,
    \end{align}
    with solutions
\begin{align}\label{eq:4D27}
\lim_{z\to1}\tilde{Z}(z)&= \psi_1 J_0\!\left( b (1-z) \right) 
+ \psi_2 Y_0\!\left( b (1-z) \right)
\nonumber\\ 
&\approx
\psi_1 + \psi_2 \left( \frac{2}{\pi} \gamma + \frac{2}{\pi} \log \big(b (1-z)\big) \right).
    \end{align}
 Again, to avoid divergences, we impose $\psi_2=0$.

    \item \textbf{Case 3:} $\frac{\lambda^2}{\alpha^2}(1+2h\kbar) > 2$.  
    In this regime, the equation becomes
    \begin{align}
        \tilde{Z}''(z)&-\frac{1}{1-z}\tilde{Z}'(z)-\frac{a^2}{1-z}\tilde{Z}(z)=0,
    \label{eq:4D28}\\
    a^2&=\frac{\frac{\lambda^2}{\alpha^2}(1+2h\kbar)-2}{3-h^2-q^2}>0,
    \end{align}
    whose solutions are 
    \begin{align}\label{eq:4D29}
    \lim_{z\to1}\tilde{Z}(z)&= \psi_1 I_0\!\left( 2a \sqrt{1-z} \right) 
+ \psi_2 K_0\!\left( 2a \sqrt{1-z} \right)
\nonumber\\ 
&\approx
\psi_1 + \psi_2 \left(- \gamma -\log \big(a (1-z)\big) \right).
    \end{align}
To ensure regularity, we again choose $\psi_2=0$ as the boundary condition.
\end{itemize}

\subsection{Numerical method and results}
\label{subsec:num_method_results}

\begin{figure}[!t]
  \centering
  \includegraphics[width=1.0\linewidth]{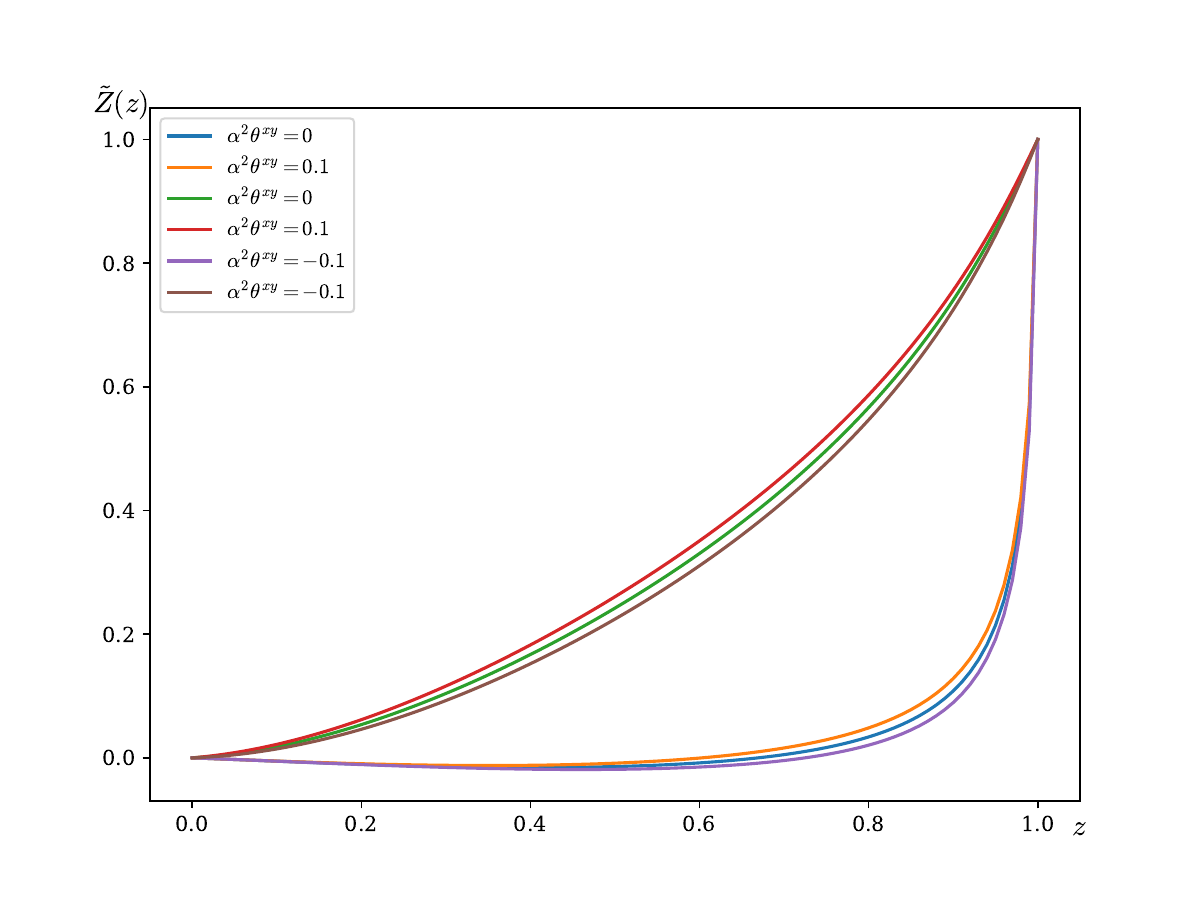}
  \caption{Zero- and one-node modes of the scalar field with $\mathcal{O}_2$ turned on, shown for both the NC case and the commutative case. The magnetic charge parameter is fixed at $h=0.5$.}
  \label{fig:two_possible_solutions}
\end{figure}

Equation \eqref{eq:4D21} cannot be solved analytically, and therefore we have to employ numerical methods. In our analysis, we use the shooting method with initial conditions
\begin{equation}
\tilde{Z}(1)=1, 
\quad
\tilde{Z}'(1)=\frac{2-\frac{\lambda^2}{\alpha^2}(1+2h\kbar)}{3-h^2-q^2} \,.
    \label{eq:4D30}
\end{equation}
The system depends on several parameters: $\frac{\lambda^2}{\alpha^2},h,q,\kbar$. The value of $\frac{\lambda^2}{\alpha^2}$ was previously chosen (taking $n=0$) to be $\frac{\lambda^2}{\alpha^2}=\frac{B}{\alpha^2}=2h$. This leaves us with three free parameters: $\kbar$, $h$, and $q$. However, for arbitrary values of these parameters, solutions do not exhibit the correct asymptotic behavior compatible with \eqref{eq:4D22}. Consequently, in our method, for a fixed value of $\kbar$, we fix the value of $h$ and then vary $q$ until the boundary conditions at the conformal boundary are satisfied. We also have to bear in mind that our first-order perturbative analysis demands $|h\kbar|\ll 1$ and take this condition into account in the numerical analysis.  
There are two classes of solutions: zero- and one-node modes, shown in Figure \ref{fig:two_possible_solutions}. Nevertheless, as argued in Ref.~\cite{Gubser:2008px}, only zero-node modes are physically relevant.
This proves that there exists a critical magnetic field above which condensation is not possible, in which case the trivial solution for the scalar field is preferred.
For convenience, we define the rescaled temperature
\begin{equation}
\tilde{T} = \alpha \frac{3-\tilde{q}^2}{4\pi},
\label{eq:4D31}
\end{equation}
corresponding to the temperature at zero magnetic field at which nontrivial solutions appear, with the corresponding critical value $\tilde{q}$ of the charge. An important dimensionless quantity of interest is 
\begin{equation}
\frac{B}{\tilde{T}^2} = \Bigg(\frac{4\pi}{3-\tilde{q}^2}\Bigg)^2 2h .
    \label{eq:4D32}
\end{equation}
We therefore study its dependence on $T/\tilde{T}$ and compare it to the commutative case reported in ~\cite{Albash:2008bm}.
\begin{figure}[!t]
    \centering
  \includegraphics[width=1.0\linewidth]{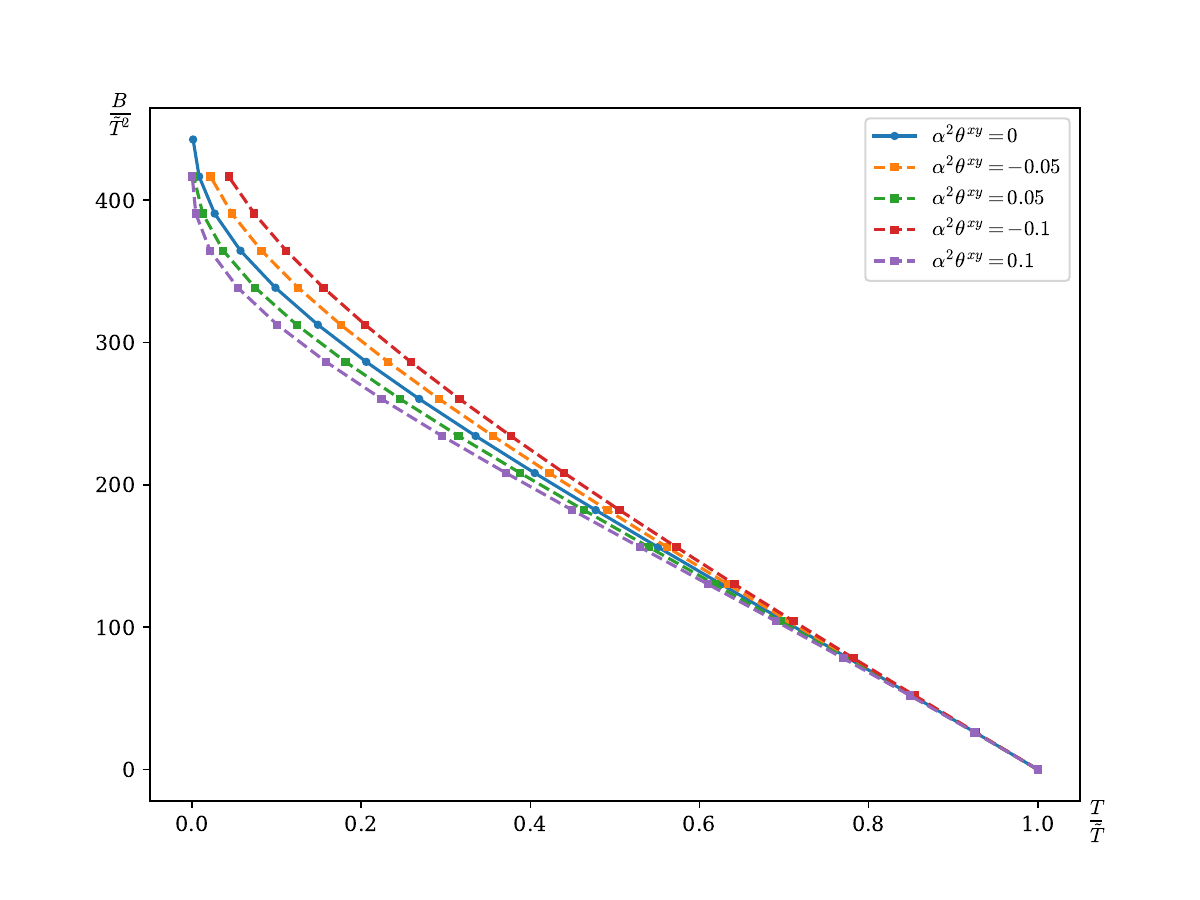}
  \caption{Allowed values of the magnetic field $B$ and temperature $T$ for condensation to occur, shown for both the NC case and the commutative case. Each curve gives us the critical value of the magnetic field for a given temperature. The region under a curve is the condensation region.}
  \label{fig:critical_B_on_T}
\end{figure}

As indicated by the numerical results in Fig. \ref{fig:critical_B_on_T}, by varying the NC parameter $\kbar$, we can continuously deform the commutative curve ($\alpha^{2}\theta^{xy}=\kbar=0$). Note that, since the analysis is limited to the first-order NC correction, the results  depend on the sign of $\kbar$. For positive/negative $\kbar$, the value of the magnetic field under which condensation is possible decreases/increases, i.e., the external magnetic field effectively increases/decreases. In addition to that, the difference between the NC case and the commutative case disappears for $B=0$, which is expected, since the NC parameter $\kbar$ always couples directly to the magnetic charge $h$.

\begin{figure}[!t]\centering\includegraphics[width=1.0\linewidth]{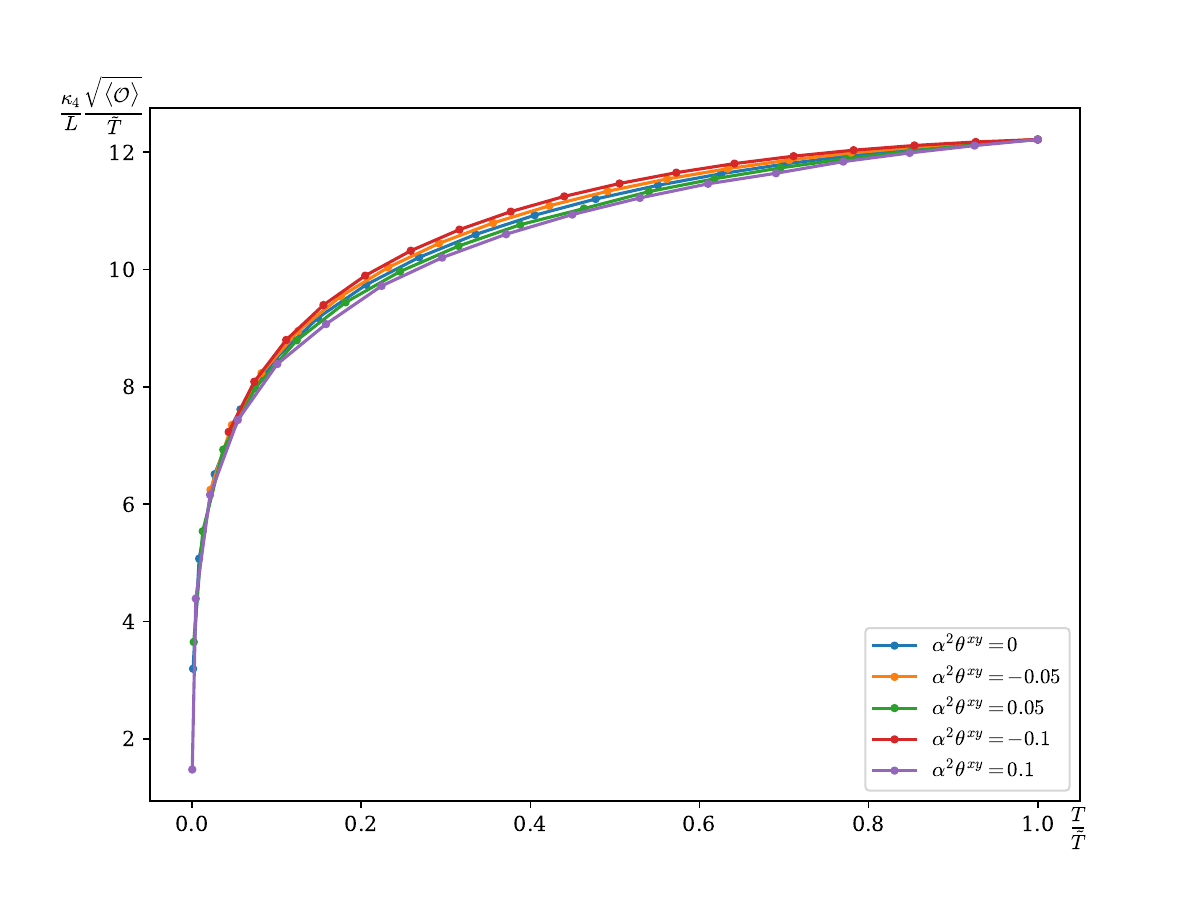}\caption{Numerical calculation of the $\langle\mathcal{O}_{2}\rangle$   temperature dependence  for a particular horizon normalization, $\widetilde Z(1)=1$. Since the value of the VEV depends on the normalization choice, the obtained results do not represent the physical amplitude of the condensate. However, the results do indicate a continuous deformation due to NC twist.}\label{fig:vev_on_T}\end{figure}

Another quantity relevant for the characterization of the condensate is the VEV of the conformal dimension $\Delta = 2$ operator $\mathcal{O}_{2}$, which is related to the bulk field expansion coefficient $\Psi^{(2)}$ through the holographic dictionary. However, in the presence of bulk noncommutativity, the holographic dictionary itself becomes deformed. A detailed analysis of this deformation is presented in Appendix \ref{appendixC}. Here, we simply state the final result  
\begin{align} \label{eq:4D33}
\frac{\kappa_4 \, \sqrt{ \langle \mathcal{O}_2} \rangle}{L\tilde{T}}
&= \frac{4\pi}{3 - \tilde{q}^2} \, \sqrt{ \frac{ \Psi^{(2)}(1-h\kbar )}{2}}\nonumber\\
&\approx \frac{4\pi}{3- \tilde{q}^2}\sqrt{\frac{\Psi^{(2)}}{2}}\bigg(1-
\frac{1}{2}h\kbar\bigg). 
\end{align}
As in~\cite{Albash:2008bm}, we  emphasize that by treating the gauge field as part of the fixed background, we neglect the nonlinear effects due to scalar backreaction. This leaves us with a linear homogeneous differential equation for the zero mode $\widetilde Z(z)$. Consequently, every solution, and therefore the coefficient $\Psi^{(2)}$, is determined only up to an overall multiplicative constant. In our numerical analysis, this constant is fixed by choosing horizon normalization
$
\widetilde Z(1)=1.
$
However, this is a numerical convention and does not determine the physical value of the condensate amplitude given by the boundary VEV $\langle\mathcal{O}_{2}\rangle$. Different choices of normalization lead to the same onset curve $B_c(T,\kbar)$, but they rescale $\Psi^{(2)}$, and hence, by the holographic dictionary, the VEV. A dynamical determination of the condensate amplitude would require solving the coupled nonlinear Maxwell--scalar-field equations, or performing a controlled expansion in which the scalar backreacts on the gauge field. Nevertheless, the NC twist deformation of the holographic dictionary remains valid even if we ignore the scalar backreaction. The numerical results shown in Fig.~\ref{fig:vev_on_T} should therefore be interpreted as illustrating the temperature dependence of the horizon-normalized VEV, together with its continuous NC deformation, rather than as determining the physical amplitude of the condensate.

\section{4D Twisted Holographic Superconductor in an External Magnetic Field}
\label{sec:4D}

In the previous section, we analyzed the effects of an Abelian Killing twist deformation of the bulk on the parameters of the $(2+1)$-dimensional holographic superconductor in an external magnetic field. Such an analysis can, to a certain extent, be extended to higher-dimensional cases. Here, we focus on the $(3+1)$-dimensional case. The topic of holographic superconductors in $(3+1)$ dimensions, both with and without an external magnetic field, has been extensively investigated \cite{Gregory:2009fj,Pan:2009xa,Pan:2010px,Liu:2010ka,Ge:2010aa}. Since in a five-dimensional bulk the Gauss-Bonnet term is no longer topological, most of these studies include it. However, in our case, the twist effects are of primary interest. As we work with Killing twists, where the background metric remains undeformed, we restrict ourselves to the pure Einstein-Hilbert gravitational action. Therefore, our five-dimensional bulk action is
\begin{multline}
S_{\text{NC}} = \frac{1}{2\kappa_5^2} \int d^5 x \, \sqrt{-g}  \Bigg(
R + \frac{12}{L^2} \\
+ L^2 \Bigg[
- \frac{1}{4} g^{\mu \alpha}  g^{\nu \beta}  \hat{F}_{\mu \nu} \star \hat{F}_{\alpha \beta} \\
- g^{\mu \nu}  (D_\mu \hat{\Phi})^* \star D_\nu \hat{\Phi} 
- m^2 \hat{\Phi}^* \star \hat{\Phi} 
\Bigg] 
\Bigg).
\label{eq:5D3}
\end{multline}
Since in five dimensions the electric and magnetic fields are no longer dual to each other, the construction of dyonic black branes with AdS asymptotics becomes considerably more involved. Therefore, unlike the four-dimensional case, where the gravitational and gauge fields constitute a fixed background, here both the scalar and the gauge field are considered dynamical. Consequently, the AdS planar black brane is taken as a fixed geometrical background with the metric
\begin{equation}
ds^2 =   -f(r) dt^2 + \frac{r^2}{L^2}(dx^2 + dy^2 + dw^2)  + \frac{1}{f(r)}dr^2,
\end{equation}
where $f(r)=\frac{r^2}{L^2}(1-\left(\frac{r_h}{r}\right)^4)$.The temperature of this black brane is given by
$
T = \frac{r_h}{\pi L^2}.
$
For subsequent calculations, it is convenient to introduce the Poincar\'e coordinate
$
z = \frac{r_h}{r},
$
in terms of which the metric takes the form
\begin{equation}
ds^2 = \frac{L^2 \alpha^2}{z^2} \left( -f(z)\, dt^2 + dx^2 + dy^2 + dw^2 \right) 
+ \frac{L^2}{z^2} \frac{dz^2}{f(z)},
\label{eq:5D2}
\end{equation}
where $\alpha = \frac{r_h}{L^2}$ and $f(z) = 1 - z^4$. 
Note that the temperature can now be written as
$
T = \frac{\alpha}{\pi}.
$
Among the generators of the $\mathbb{R} \oplus ISO(3)$ Killing algebra of the background spacetime we again consider $\partial_x$ and $\partial_y$,
\begin{equation}
\mathcal{F} = \exp\left[-i\theta^{xy}\left(\partial_x \otimes \partial_y - \partial_y \otimes \partial_x\right)\right].
\end{equation}
Using the same procedure as in the previous case, we obtain exactly the same first-order NC correction as \eqref{eq:4D12}, together with the equations of motion, only now in five dimensions. 

Following the logic of \cite{Abrikosov1963,Ge:2010aa} we will further consider separately strong and weak magnetic field limits. In the weak magnetic field limit there are no NC corrections and our analysis reduces completely to that of \cite{Ge:2010aa,Gregory:2009fj} (for $D=5$ and in the pure Einstein-Hilbert case). Therefore, we focus on the strong magnetic field limit, where following \cite{Abrikosov1963} we assume
\[
A_t = A_t(z), \quad A_x = -\frac{B}{2}y, \quad A_y = \frac{B}{2}x, \quad A_z = A_w = 0,
\]
and $\Phi = \Phi(x,y,z)$. We choose the axially symmetric gauge, since we are interested in axially symmetric configurations. It is therefore convenient to switch to polar coordinates $(x,y)\to(r,\varphi)$, implying $\Phi = \Phi(r,\varphi,z) = \Phi(r,z)$. 
Under these assumptions, the equations of motion for $A_t$ and $\Phi$ reduce to
\begin{equation}
A_t''(z) - \frac{1}{z} A_t'(z) 
- \frac{2 L^2 A_t \Phi}{f(z)z^2}
\left(\Phi - \frac{B \theta^{xy}}{2} r \frac{\partial \Phi}{\partial r}\right) = 0,
\label{eq:5D9}
\end{equation}

\begin{align}\label{eq:5D10}
z^2 f'(z) \frac{\partial \Phi}{\partial z}
- z f(z) \left(3 \frac{\partial \Phi}{\partial z} - z \frac{\partial^2 \Phi}{\partial z^2} \right) \nonumber\\
+ \frac{z^2}{\alpha^2}(1+B\theta^{xy})\left( \frac{\partial^2 \Phi}{\partial r^2}  + \frac{1}{r}\frac{\partial \Phi}{\partial r}\right)
+ \frac{z^2 A_t^2(z)}{\alpha^2 f(z)}\Phi \nonumber\\
- m^2 L^2 \Phi 
- \frac{B^2 z^2 r^2}{4 \alpha^2}(1+B\theta^{xy})\Phi = 0.
\end{align}
The scalar field equation can be solved by separation of variables, $\Phi(r,z)=R(r)Z(z)$, which transforms the partial differential equation \eqref{eq:5D10} into a system of two coupled ordinary differential equations
\begin{equation}
R''(r) + \frac{1}{r} R'(r) - \Big(\frac{Br}{2}\Big)^2 R(r) = -\lambda^2 R(r),
\label{eq:5D11}
\end{equation}
\begin{align}\label{eq:5D12}
Z''(z) &+ \Bigg(\frac{f'(z)}{f(z)} - \frac{3}{z} \Bigg) Z'(z) 
+ \Bigg( \frac{A_t^2(z)}{\alpha^2 f^{2}(z)} \nonumber\\
-& \frac{m^2 L^2}{z^2 f(z)} 
- \frac{\lambda^2}{\alpha^2 f(z)}(1+ B \theta^{xy}) \Bigg) Z(z) = 0.
\end{align}
Equation \eqref{eq:5D12}, similar to \eqref{eq:4D17}, admits exact solutions in terms of hypergeometric functions. However, for $\lambda^2 = B (2n+1)$, with $n \in \mathbb{N}_0$, it reduces to the equation of a two-dimensional harmonic oscillator. Since the lowest modes are expected to condense first, we again choose $n=0$, which implies
\begin{equation}
R(r)=\exp\left(-\frac{Br^2}{4}\right).
\label{eq:5D13}
\end{equation}
The radial distribution of the condensate is the same as in the $(2+1)$-dimensional case, indicating strong localization and the formation of superfluid droplets.

After substituting $\lambda^2=B$ and fixing $m^2L^2=-3$, which implies $\Delta=3$ or $1$, Eq. \eqref{eq:5D12} reduces to
\begin{align}\label{eq:5D14}
Z''(z) &+ \bigg(\frac{f'(z)}{f(z)} - \frac{3}{z} \bigg) Z'(z) 
+ \bigg( \frac{A_t^2(z)}{\alpha^2 f(z)^2} \nonumber\\
&+ \frac{3}{z^2 f(z)} 
- \frac{B}{\alpha^2 f(z)}(1+ B \theta^{xy}) \bigg) Z(z) = 0.
\end{align}
In order to solve the coupled system of equations \eqref{eq:5D9} and \eqref{eq:5D14}, we employ two approaches -- one analytical and one numerical —- both based on an appropriate approximation. We also have to take into account the perturbative regime we work in, $ |B\theta^{xy}|\ll 1$.

Since we are in the vicinity of the critical point, the scalar field can be considered negligibly small compared to the gauge field. Therefore, neglecting it in \eqref{eq:5D9} renders an analytically solvable equation. Its solution, subject to the appropriate boundary conditions, is given by
\begin{equation}
A_t(z)=\frac{\rho}{r_h^2}(1-z^2).
\end{equation}
Now we tackle the equation for $Z(z)$ using analytical and numerical methods. 

\subsection{Analytical method}
\label{subsec:analytical}
Here we employ the analytical method developed in 
\cite{Gregory:2009fj,Pan:2009xa,Pan:2010px,Liu:2010ka,Ge:2010aa}, 
to determine the temperature dependence of the critical magnetic field. First, we expand $Z(z)$ both near the asymptotic boundary ($z=0$), 
\begin{align}
Z(z) = c_1 z + c_3 z^3 + \mathcal{O}(z^4),
\end{align}
and near the horizon ($z=1$),
\begin{align}
Z(z) &= Z(1) - Z'(1)(1-z) \nonumber\\
&+ \frac{1}{2} Z''(1)(1-z)^2 + \mathcal{O}((1-z)^3).
\label{eq:5D15}
\end{align}
Substituting \eqref{eq:5D15} in Eq. \eqref{eq:5D14} and expanding around the horizon yields
\begin{equation}
Z'(1) = \frac{Z(1)}{4}\left(3-\frac{B}{\alpha^2}\left(1+B\theta^{xy}\right)\right),
\end{equation}
together with
\begin{equation}
\begin{split}
Z''(1)=&-\frac{Z(1)}{32}\Bigg(
15+\frac{6B}{\alpha^2}
-\frac{2B^3\theta^{xy}}{\alpha^4}
-\frac{B^4(\theta^{xy})^2}{\alpha^4}
\\
&-\frac{B^2}{\alpha^4}
+\frac{6B^2\theta^{xy}}{\alpha^2}
+\frac{A_t'(1)^2}{\alpha^2}
\Bigg).
\end{split}
\end{equation}
Thus, the coefficients in the near-horizon expansion are fully determined by $Z(1)$. Near the asymptotic boundary, we impose $c_1=0$, which corresponds to the standard quantization procedure.

The central idea of this approximate analytical method is to smoothly match the near-horizon and near-boundary expansions at an intermediate point, chosen as $z_m=\frac{1}{2}$. This leads to a system of equations relating $Z(1)$, $c_3$, and $A_t'(1)$. Eliminating $Z(1)$ and $c_3$, we obtain
\begin{equation}\label{eq:5D16}
\begin{split}
&309+\frac{98B}{\alpha^2}
+\frac{10B^3\theta^{xy}}{\alpha^4}
+\frac{5B^4(\theta^{xy})^2}{\alpha^4}
\\
&+\frac{5B^2}{\alpha^4}
+\frac{98B^2\theta^{xy}}{\alpha^2}
-\frac{5A_t'(1)^2}{\alpha^2}=0.
\end{split}
\end{equation}
Solving for $B$, taking the positive root, and expanding to the first order in $\theta^{xy}$, we find the critical magnetic field,
\begin{align}\label{eq:5D17}
&B_c=\frac{1}{5}
\left(
-49 \alpha^{2}
+ \sqrt{
\alpha^{2}
\left(
856 \alpha^{2}
+ 25 \left(A_t'(1)\right)^{2}
\right)
}
\right)\nonumber\\
&-
\frac{\theta^{xy}}{25}
\left(
-49 \alpha^{2}
+ \sqrt{
\alpha^{2}
\left(
856 \alpha^{2}
+ 25 \left(A_t'(1)\right)^{2}
\right)
}
\right)^{2}.
\end{align}
Using $\alpha=\pi T$, $A_t'(1)=\frac{2\rho}{r_h^2}$, and the critical temperature obtained from the weak magnetic field analysis conducted in \cite{Gregory:2009fj}, namely,
\begin{equation}
T_c=\left(\frac{2\rho}{L}\sqrt{\frac{5}{309}}\right)^{\frac{1}{3}}\frac{1}{\pi L},
\end{equation}
we arrive at the temperature dependence of the critical magnetic field including the NC correction,
\begin{align}\label{eq:5D18}
&B_c=\frac{\pi^{2}T^2}{5}
\left(
-49 + \sqrt{856 + 1545 \left(\frac{T_c}{T}\right)^{6}}
\right)\nonumber\\
&-
\left[
\frac{\pi^{2}T^2}{5}
\left(
-49 + \sqrt{856 + 1545 \left(\frac{T_c}{T}\right)^{6}}
\right)
\right]^{2}
\theta^{xy}.
\end{align}
Note that this result can be written as
\begin{equation}
B_c = B_c^{(0)}\left(1 - B_c^{(0)}\theta^{xy}\right),
\label{eq:5D19}
\end{equation}
where
\begin{equation}
B_c^{(0)}=\frac{\pi^{2} T^{2}}{5}
\left(
-49 + \sqrt{856 + 1545 \left(\frac{T_c}{T}\right)^{6}}
\right)
\label{eq:5D20}
\end{equation}
represents the temperature dependence of the critical magnetic field in the undeformed theory. 

\begin{figure}[!t]
  \centering
\includegraphics[width=1.0\linewidth]{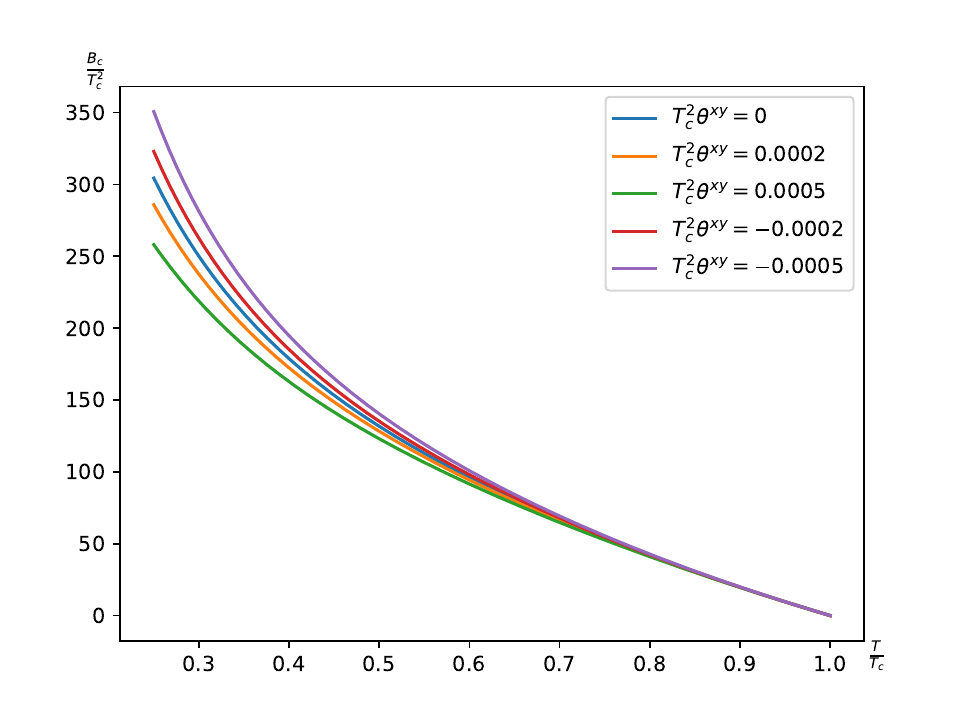}
  \caption{Analytical results for the temperature dependence of the critical magnetic field for different values of the NC parameter indicate a continuous deformation of the critical curve.}
  \vspace{-10pt}
  \label{fig:5D_analytic}
\end{figure}

The classical result for $B_c^{(0)}$ coincides exactly with the result obtained in \cite{Ge:2010aa} in the pure Einstein-Hilbert limit (i.e., in the absence of the Gauss-Bonnet term), which confirms the consistency of our analysis. It is more convenient to plot a dimensionless quantity $B_c/T_c^2$ and to take $T_c^2\theta^{xy}$ as a dimensionless deformation parameter. Figure \ref{fig:5D_analytic} shows us that, for positive/negative values of the NC deformation parameter $T_c^2\theta^{xy}$, the critical magnetic field decreases/increases, implying that the NC twist effectively increases/decreases the  magnetic field. It is interesting to note that \eqref{eq:5D19} implies that the magnetic field is effectively modified just like the NC Landau levels
$B_{\text{eff}} = B(1 + B\theta)$ from \cite{DimitrijevicCiric:2018so23NCED}.

\subsection{Numerical method}

\begin{figure}[!t]
  \centering
 \includegraphics[width=1.0\linewidth]{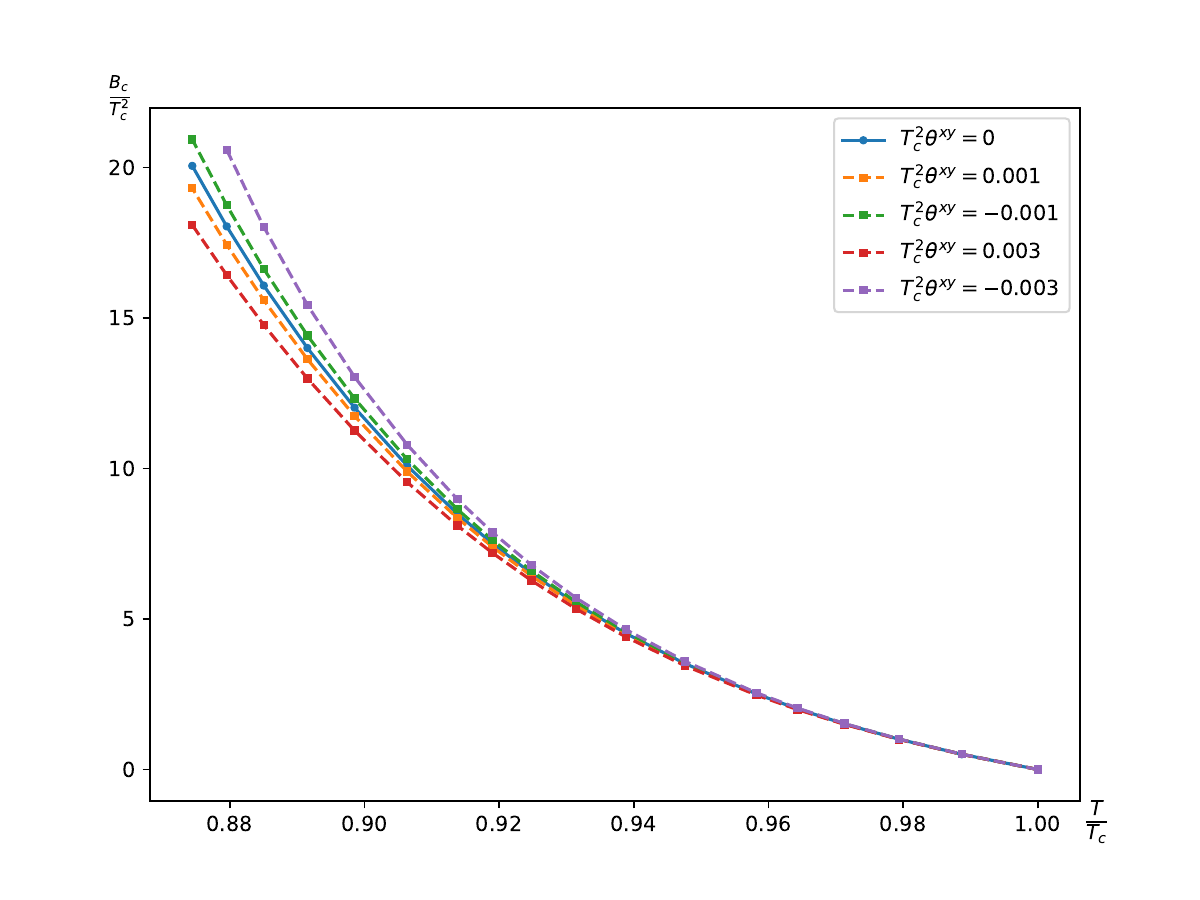}
  \caption{Numerical results consistently indicate a continuous deformation of the critical curve due to NC twist.}
  \label{fig:5D_numeric}
\end{figure}

After the Eq. \eqref{eq:5D9} is solved analytically, a numerical method similar to the one applied in the $(2+1)$-dimensional case can also be used to solve \eqref{eq:5D12}. Unlike the analytical method presented in the previous subsection, which was approximate, the numerical approach based on the shooting method does not assume any additional approximations.
The results of the numerical calculation are shown in Fig. \ref{fig:5D_numeric}. Note that there is qualitative agreement with the analytical result. The numerical calculation is stable only in the vicinity of the critical temperature, specifically for $\frac{T}{T_c}\in[0.87,1]$, and we had to take larger values of the NC parameter compared to the analytical case to make the deformation of the critical curve visible.  

\section{Conclusion and Outlook}
\label{sec:conclusions}

In this paper, we initiated a systematic study of the role of NC gauge field theory in a holographic description of condensed matter systems, in particular, holographic superconductors. We focused on $(2+1)$- and $(3+1)$-dimensional models of holographic superconductors (or, more accurately, charged superfluids) with established gravitational bulk actions. Working in the probe limit, we applied a NC twist deformation of the bulk fields and promoted the classical bulk action into a NC action invariant under deformed gauge transformations. We used a special kind of twist adapted to the isometries of the background black hole  geometry that only affects the gauge-matter sector, leaving the metric intact. The goal was to see how this Killing twist transformation in the bulk is reflected in the dual boundary system through condensation parameters such as the critical magnetic field. Using mainly numerical methods for solving the relevant field equations, we showed that critical behavior indeed depends on the parameter of noncommutativity. As the obtained results clearly indicate, by varying the NC parameter, we can continuously change the value of the critical magnetic field for a given temperature. The NC twist also deforms the holographic dictionary.    

It should be noted that, in this setup, the boundary system is to be regarded as a model of some potentially realistic superconductor, while the gravitational dual amounts to just an alternative formal description. In that sense, the use of a NC field theory is simply an expansion of the holographic vocabulary, providing additional structure that might capture some relevant aspects of the physics of superconductors and related condensed matter theory (CMT) systems in the holographic setting. From this point of view, NC gauge field theory need not be regarded as a physical theory, merely as a part of the holographic language, augmenting the dual gravitational description of lower-dimensional strongly correlated systems. 
In this paper, we used a very simple class of twist deformations that introduces only NC gauge and matter fields, but does not affect the bulk geometry. This was only a first step, setting the stage for a more general investigation of the full capacity of twist deformations as a tool in AdS/CMT. Further investigation could involve bulk geometries with NC-deformed metric, and even going into nonperturbative regime of NC field theory/gravity. 
An ultimate goal is to identify a concrete physical aspect of some CMT system that has a characteristic dual description in terms of NC bulk fields, including the metric.  

Apart from these general considerations regarding the role of NC field theory, we also point out that both bulk models we discussed assume the probe limit. In the $(2+1)$-dimensional model, the metric and the gauge field together constitute a self-consistent Einstein-Maxwell background on which the scalar propagates. In the $(3+1)$-dimensional bulk model, only geometry is fixed while the gauge field and the scalar are both dynamical. Holographic models with dynamical gravity have been studied in the context of AdS/CMT, but the addition of a NC deformation presents a highly nontrivial complication. This interesting but challenging task demands systematic further investigation.

Finally, let us mention that, although there are several papers concerning holographic superconductors that introduce a NC scale to modify the bulk \cite{Pramanik:2014,Ghorai:2016,Pramanik:2015,Gangopadhyay:2018,Paul:2025}, their approach is essentially phenomenological, based on smearing the sources (mass and charge distributions of a black hole) with the NC scale as a kind of screening parameter  
and not on a fundamental introduction of a NC-deformed algebra at the field-theory level. However, it is interesting to note that \cite{Pramanik:2015} predicts decreases in the critical magnetic field due to the presence of the NC scale. Our results, on the other hand, indicate a continuous deformation in both directions of the critical curve, controlled by the strength of the NC twist. 

\section*{Acknowledgement}
\label{sec:acknowledgement}
We thank Du\v{s}an \Dj or\dj evi\'{c} and Aleksandra Go\v{c}anin for many helpful discussions and suggestions. This work is supported by the funding provided by the Faculty of Physics, University of Belgrade, through Grant No. 451-03-136/2025-03/200162 by the Ministry of Science, Technological Development and Innovations of the Republic of Serbia. D.G. acknowledges the support of the Science Fund
of the Republic of Serbia, Grant No. 9029-YF-SAIGE, Twisted Holography: A Holographic Stance on the Quantum Superposition of
Spacetimes - HOLISTIQUS.

\appendix

\section{NC expansions}
\label{appendixA}

The general rule for expanding the product of two NC fields, $\hat{X}=X+\hat{X}^{(1)}+\mathcal{O}(\theta^{2})$ and $\hat{Y}=Y+\hat{Y}^{(1)}+\mathcal{O}(\theta^{2})$, up to first order reads
\begin{equation}\label{A1}
(\hat{X}\star\hat{Y})^{(1)}=\hat{X}^{(1)}Y+X\hat{Y}^{(1)}+\frac{i}{2}\theta^{\alpha\beta}\partial_{\alpha}X\partial_{\beta}Y,    
\end{equation}
where the last term comes from the expansion of the $\star$-product. Now, according to the representation in which they transform, NC fields have a specific structure of their first-order terms. For a NC field $\hat{\Psi}$ from the fundamental representation we have
\begin{equation}\label{A2}
\hat{\Psi}^{(1)}=-\frac{1}{4}\theta^{\alpha\beta}A_{\alpha}(\partial_{\beta}+D_{\beta})\Psi+\text{Cov}(\hat{\Psi}^{(1)}),
\end{equation}
where the Cov term is built out of covariant blocks, while the first term explicitly involves the gauge field and the partial derivative. Using the general rule \eqref{A1}, it can be easily shown that for a pair of such fields we have
\begin{align}\label{A3}
(\hat{\Psi}_{1}\star\hat{\Psi}_{2})^{(1)}=   &-\frac{1}{4}\theta^{\alpha\beta}A_{\alpha}(\partial_{\beta}+D_{\beta})(\Psi_{1}\Psi_{2}) \nonumber\\
&+\frac{i}{2}\theta^{\alpha\beta}D_{\alpha}\Psi_{1}D_{\beta}\Psi_{2}\nonumber\\
&+\text{Cov}(\hat{\Psi}^{(1)}_{1})\Psi_{2}+\Psi_{1}\text{Cov}(\hat{\Psi}_{2}^{(1)}).
\end{align}
On the other hand, for a field transforming in the adjoint representation we have
\begin{equation}
\hat{\Xi}^{(1)}=-\frac{1}{4}\theta^{\alpha\beta}\{A_{\alpha},(\partial_{\beta}+D_{\beta})\Xi\}+\text{Cov}(\hat{\Xi}^{(1)}),    
\end{equation}
and for the product of two adjoint fields
\begin{align}\label{A5}
(\hat{\Xi}_{1}\star\hat{\Xi}_{2})^{(1)}=   &-\frac{1}{4}\theta^{\alpha\beta}\{A_{\alpha},(\partial_{\beta}+D_{\beta})(\Xi_{1}\Xi_{2})\} \nonumber\\
&+\frac{i}{2}\theta^{\alpha\beta}D_{\alpha}\Xi_{1}D_{\beta}\Xi_{2}\nonumber\\
&+\text{Cov}(\hat{\Xi}^{(1)}_{1})\Xi_{2}+\Xi_{1}\text{Cov}(\hat{\Xi}_{2}^{(1)}),    
\end{align}
where $D_{\mu}\Xi=\partial_{\mu}\Xi-i[A_{\mu},\Xi]$ for a general gauge group.  

Consider a complex scalar field $\hat{\Phi}$ transforming in the fundamental representation. Its first-order correction does not have the Cov term and is simply given by
\begin{align}
\hat{\Phi}^{(1)}&=-\frac{1}{4}\theta^{\alpha\beta}A_{\alpha}(\partial_{\beta}+D_{\beta})\Phi.
\end{align}
Its covariant derivative $D_{\mu}\hat{\Phi}=\partial_{\mu}\hat{\Phi}-i\hat{A}_{\mu}\star\hat{\Phi}$ also transforms in the fundamental representation and has the expansion
\begin{align}
(D_{\mu}\hat{\Phi})^{(1)}=&-\frac{1}{4}\theta^{\alpha\beta}A_{\alpha}(\partial_{\beta}+D_{\beta})D_{\mu}\Phi\nonumber\\
&+\frac{1}{2}\theta^{\alpha\beta}F_{\alpha\mu}D_\beta\Phi.    
\end{align}
On the other hand, the field strength $\hat{F}_{\mu\nu}$ is an adjoint field, and its first-order correction is
\begin{align}
\hat{F}^{(1)}_{\mu\nu} =& - \frac{1}{4} \, \theta^{\alpha\beta} \{ A_\alpha, (\partial_\beta + D_\beta) F_{\mu\nu} \} \nonumber\\
&+ \frac{1}{2} \, \theta^{\alpha\beta} \{ F_{\alpha\mu}, F_{\beta\nu} \}.
\end{align}
Now, we can calculate the first-order corrections of the products $\hat{\Phi}^{*}\star\hat{\Phi}$, $D_{\mu}\hat{\Phi}^{*}\star  D_{\nu}\hat{\Phi}$ and $\hat{F}_{\mu\nu}\star\hat{F}_{\rho\sigma}$ that appear in the NC action (\ref{eq:4D10}). 
For that matter, it is very important that these products appear under the integral and that we use an Abelian Killing twist that leaves the measure invariant. Up to first order in the twist parameter \(\theta\), the $\star$-product of two fields is given by
\begin{equation}
f_{1} \star f_{2}
=
f_{1}f_{2}
+
\frac{i}{2}\theta^{IJ}X_I [f_{1}]X_J[f_{2}]
+
O(\theta^2).
\end{equation}
For a Killing twist, with $X_{I}=K_{I}$, the measure $\sqrt{-g}$ couples classically (no $\star$-product),
\begin{align}
\int d^4x\,\sqrt{-g}\, (f_{1} \star &g_{2})
=
\int d^4x\,\sqrt{-g}\, f_{1} f_{2}
\\
&+
\frac{i}{2}\theta^{IJ}
\int d^4x\,\sqrt{-g}\, K_I [f_{1}]K_J [f_{2}].\nonumber
\end{align}
Using integration by parts with respect to \(K_I\), noting that $\mathcal{L}_{K_{I}}[g_{\mu\nu}]=0$, and assuming boundary terms vanish, one finds
\begin{equation}
\int d^4x\sqrt{-g}\, K_I [f_{1}]K_J [f_{2}]
=
-\int d^4x\sqrt{-g}\, f_{1}\, K_I [K_J [f_{2}]] .
\end{equation}
Since the twist is Abelian, the vector fields commute $[K_I,K_J]=0$,
so $\theta^{IJ} K_I K_J = 0$. 
It follows that, up to boundary terms, the first-order correction due to $\star$-product vanishes,
\begin{equation}
\int d^4x\,\sqrt{-g}\, (f_{1} \star f_{2})
=
\int d^4x\,\sqrt{-g}\, f_{1} f_{2}.
\end{equation}
However, if the fields $f_{1}$ and $f_{2}$ are NC fields, then we still have to take into account their SW $\theta$-expansion. Therefore, we have
\begin{align}
&\int d^4x\,\sqrt{-g}\, (\hat{f}_{1} \star \hat{f}_{2})
=
\int d^4x\,\sqrt{-g}\, \hat{f}_{1} \hat{f}_{2}\\
&=\int d^4x\,\sqrt{-g}\Big(f_{1}f_{2}+\hat{f}^{(1)}_{1}f_{2}+f_{1}\hat{f}^{(1)}_{2}\Big)+\mathcal{O}(\theta^{2}).\nonumber
\end{align}
We see that the first-order NC correction exists, but not due to the $\star$-product between the fields (which reduces to the ordinary product for an Abelian Killing twist) but because of the $\theta$-expansion of the NC fields themselves. This is the situation with our NC action. In a more general case, we would have to use Eqs. \eqref{A3} and \eqref{A5} that include the contribution of the $\star$ to the product of NC fields. 

For the NC mass term, we thus have
\begin{align}
-&m^{2}\int d^{4}x\sqrt{-g}\;(\hat{\Phi}^{*}\star\hat{\Phi})^{(1)}=-m^{2}\int d^{4}x\sqrt{-g}(\hat{\Phi}^{*}\hat{\Phi})^{(1)}\nonumber\\
&=-m^{2}\int d^{4}x\sqrt{-g}(\hat{\Phi}^{*(1)}\Phi+\Phi^{*}\hat{\Phi}^{(1)})\nonumber\\
&=\frac{m^{2}}{4}\theta^{\alpha\beta}\int d^{4}x\sqrt{-g}\;A_{\alpha}(\partial_{\beta}+D_{\beta})\vert\Phi\vert^{2}\nonumber\\
&=\frac{m^{2}}{4}\theta^{\alpha\beta}\int d^{4}x\sqrt{-g}\;F_{\alpha\beta}\vert\Phi\vert^{2},
\end{align}
where we used partial integration (ignoring boundary terms) and the relation  $i[D_{\alpha},D_{\beta}]\Phi=F_{\alpha\beta}\Phi$.

Next, we have
\begin{align}
-&\int d^{4}x\sqrt{-g}\;g^{\mu\nu}(D_{\mu}\hat{\Phi}^{*}\star D_{\nu}\hat{\Phi})^{(1)}\nonumber\\
=&-\int d^{4}x\sqrt{-g}\;g^{\mu\nu}\nonumber
\theta^{\alpha\beta}\bigg(-\frac{1}{4}A_{\alpha}(\partial_{\beta}+D_{\beta})(D_{\mu}\Phi^{*}D_{\nu}\Phi)\nonumber\\
&+\frac{1}{2}F_{\alpha\mu}D_\beta\Phi^{*}D_{\nu}\Phi+\frac{1}{2}D_{\mu}\Phi^{*}F_{\alpha\nu}D_\beta\Phi
\bigg)\nonumber\\
&=\int d^{4}x\sqrt{-g}\;g^{\mu\nu}\theta^{\alpha\beta}\bigg(\frac{1}{4}F_{\alpha\beta}D_{\mu}\Phi^{*}D_{\nu}\Phi)\nonumber\\
&-\frac{1}{2}F_{\alpha\mu}D_\beta\Phi^{*}D_{\nu}\Phi-\frac{1}{2}D_{\mu}\Phi^{*}F_{\alpha\nu}D_\beta\Phi
\bigg).
\end{align}
Finally, we have
\begin{align}
- &\frac{1}{4}\int d^{4}x\sqrt{-g}\; g^{\mu \rho}  g^{\nu \sigma}  (\hat{F}_{\mu \nu} \star \hat{F}_{\rho \sigma})^{(1)}\nonumber\\
=- &\frac{\theta^{\alpha\beta}}{4}\int d^{4}x\sqrt{-g}\; g^{\mu \rho}  g^{\nu \sigma}\\
\times\bigg(&- \frac{1}{4}  \{ A_\alpha, (\partial_\beta + D_\beta) F_{\mu\nu} \}F_{\rho\sigma} 
+ \frac{1}{2} \{ F_{\alpha\mu}, F_{\beta\nu} \}F_{\rho\sigma}\nonumber\\
&- \frac{1}{4}  F_{\mu\nu}\{ A_\alpha, (\partial_\beta + D_\beta) F_{\rho\sigma} \} 
+ \frac{1}{2} F_{\mu\nu} \{ F_{\alpha\rho}, F_{\beta\sigma} \}\bigg)\nonumber\\
=&\theta^{\alpha\beta}\int d^{4}x\sqrt{-g}\bigg(\frac{1}{8}F_{\alpha\beta}F_{\mu\nu}F^{\mu\nu}
-\frac{1}{2}F_{\alpha\mu}F_{\beta\nu}F^{\mu\nu}
\bigg).\nonumber
\end{align}

\section{NC gauge field equation}
\label{appendixB}

The variation of the NC action (\ref{eq:4D10}) with respect to $A_{\lambda}$ up to first order in $\theta$, yields a curved spacetime NC equation for the gauge field,  
\begin{align}\label{B1}
&\partial_\mu F^{\mu\lambda}
+ \Gamma^\rho_{\mu\rho} F^{\mu\lambda} \nonumber\\
&+ \theta^{\alpha\lambda}
\bigg(
    \frac{1}{2}
    \big(
        \partial_\mu (F_{\alpha\nu} F^{\mu\nu})
        + \Gamma^\rho_{\mu\rho} F_{\alpha\nu} F^{\mu\nu}
    \big)
    - \frac{1}{4} \partial_\alpha (F_{\mu\nu} F^{\mu\nu})
\bigg)\nonumber\\
&+ \theta^{\alpha\beta}\bigg(
    - \frac{1}{2}
    \big(
        \partial_\mu (F_{\alpha\beta} F^{\mu\lambda})
        + \Gamma^\rho_{\mu\rho} F_{\alpha\beta} F^{\mu\lambda}
    \big) \\
 &+ \partial_\mu (F_{\alpha}{}^{\mu} F_{\beta}{}^{\lambda})
    + \Gamma^\rho_{\mu\rho} F_{\alpha}{}^{\mu} F_{\beta}{}^{\lambda} 
    - \partial_\alpha (F_{\beta\mu} F^{\mu\lambda})
\bigg)
= j^\lambda,\nonumber
\end{align}
where 
\begin{align}\label{B2}
&j^\lambda=
\left(
1 - \frac{1}{4} \theta^{\alpha\beta} F_{\alpha\beta}
\right)
j^{\lambda(0)}\nonumber\\
&+\frac{1}{2} \theta^{\alpha\beta} g^{\lambda\mu}
\Big(
F_{\alpha\mu} j^{(0)}_{\beta}
+ \partial_\alpha
(
D_\mu \Phi^* D_\beta \Phi
+ D_\beta \Phi^* D_\mu \Phi
)
\Big)
\nonumber\\
&- \frac{1}{2} \theta^{\alpha\lambda}
\Big(
- F_{\alpha\nu} j^{\nu(0)}
+ g^{\mu\nu} \partial_\alpha
(
D_\mu \Phi^* D_\nu \Phi
)\nonumber\\
&\hspace{1.4cm}- \partial_\nu
\Big(
g^{\mu\nu}
(
D_\mu \Phi^* D_\alpha \Phi
+ D_\alpha \Phi^* D_\mu \Phi
)
\Big)\\\nonumber
&\hspace{1.4cm}- \Gamma^\rho_{\nu\rho}
\Big(
g^{\mu\nu}
(
D_\mu \Phi^* D_\alpha \Phi
+ D_\alpha \Phi^* D_\mu \Phi
)
\Big)
\Big),
\end{align}
is the NC source term. 
The undeformed source $j^{\lambda(0)}$ has the standard form
\begin{equation}\label{B3}
j^{\lambda(0)}
=
i g^{\mu\lambda}
\left(
D_\mu \Phi^*  \Phi
- \Phi^* D_\mu \Phi
\right).    
\end{equation}
If we set $\Phi=0$, the NC source term vanishes. Moreover, if we evaluate \eqref{B1} on our classical dyonic RN-AdS black hole background, the NC contribution drops out, meaning that this configuration remains a solution of the perturbative NC Maxwell equations, at least to the first order. Since we work in the decoupling limit in which the scalar does not backreact on the geometry and the gauge field, the dyonic RN-AdS black hole provides a consistent background for the NC scalar dynamics.

\section{Holographic renormalization}
\label{appendixC}

As shown in \eqref{eq:4D22}, the $z$-dependent part of the bulk scalar field $\Phi$, denoted by $\tilde{Z}(z)$, has the following near-boundary asymptotics:
\begin{equation}
\lim_{z\to0}\tilde{Z}(z)=\Psi^{(1)}z+\Psi^{(2)}z^2 ,
\end{equation}
with two normalizable modes, $\Psi^{(1)}$ and $\Psi^{(2)}$. We adopt the standard quantization scheme, in which $\Psi^{(1)}$ is interpreted as the source and $\Psi^{(2)}$ as the response. Thus, the full axially symmetric scalar field asymptotics reads
\begin{equation}
\lim_{z\to0}\Phi(r,z)=e^{i\varphi}\,(\phi_0(r)\,z+\chi(r)\,z^2 ),
\end{equation}
with a constant overall phase of the scalar field.  

To compute the VEV of the operator $\mathcal{O}_2$, one needs to evaluate the variation of the on-shell action with respect to the boundary scalar source. Explicitly,
\begin{multline}
\delta S_{\text{on-shell}}
= -\frac{L^2}{2 \kappa_4^2}(1-h\kbar) 
\int d^3y \, \sqrt{-g}\, g^{zz}\, \partial_z \Phi \delta \Phi^* 
\Big|_{z=0}^{z=1} \\
= \lim_{z \to 0} 
\frac{L^4\alpha^{3}}{2 \kappa_4^2}(1-h\kbar)
\int d^3y  
\Big( z^{-1} \phi_0(r)\,\delta \phi_0(r) \\
+ 2  \chi(r)\,\delta \phi_0(r) 
+  \phi_0(r)\,\delta \chi(r) 
+ \mathcal{O}(z) \Big).
\label{eq:C3}
\end{multline}
The first term in \eqref{eq:C3} diverges, so holographic renormalization must be applied by adding the counterterm 
\begin{align}
S_{\text{ct}}
&= -\frac{L}{4 \kappa_4^2}(1-h\kbar)
\int d^3y \sqrt{-\gamma} 
\vert\Phi(r, z=0)\vert^{2}\\
&= -\frac{L^{4}\alpha^{3}}{4 \kappa_4^2}(1-h\kbar)
\int d^3y \big(z^{-1}\phi^{2}_{0}(r)+2\phi_{0}(r)\chi(r)\big),\nonumber
\end{align}
where $\gamma_{ab}$ is the induced metric on the boundary, with $\sqrt{-\gamma}=\frac{L^{3}\alpha^{3}}{z^{3}}f(z)\sim\frac{L^{3}\alpha^{3}}{z^{3}}$ as $z\rightarrow 0$.  
After renormalization, the variation of the on-shell action becomes
\begin{equation}
\delta S_{\text{on-shell}}
= \frac{L^4\alpha^{3}}{2 \kappa_4^2}(1-h\kbar) 
\int d^3y  \, \chi(r) \, \delta \phi_0(r),
\end{equation}
which leads to the holographic one-point function,
\begin{equation}
\langle \mathcal{O}_2(r) \rangle
= \frac{1}{D} 
\frac{\delta S_{\text{on-shell}}}{\delta \phi_0(r)}
= \frac{L^2}{2 \kappa_4^2}(1-h\kbar)  L^2 \alpha^3  \chi(r).
\end{equation}
The factor $D$ normalizes the operator with respect to unit length in the $r$ direction. Finally, neglecting the $r$ dependence for the overall factor, we obtain
{
\setlength{\abovedisplayskip}{4pt}
\setlength{\belowdisplayskip}{4pt}
\setlength{\abovedisplayshortskip}{2pt}
\setlength{\belowdisplayshortskip}{2pt}
\begin{equation}
\langle \mathcal{O}_2 \rangle
= \frac{L^2}{2 \kappa_4^2}(1-h\kbar) \alpha^2  \Psi^{(2)}.
\end{equation}}
In this sense the holographic dictionary itself gets deformed under a NC twist. The result is valid up to the first order in the NC parameter $\kbar$, but could be formally extended to higher orders. Note again the necessity of having a magnetic field for turning on the NC effect. 

\bibliographystyle{apsrev4-2}
\bibliography{refs.bib}

\end{document}